\documentclass[aps, pra, showpacs,twocolumn,amsmath,amssymb,superscriptaddress, floatfix]{revtex4}

\usepackage{graphicx,tabularx}
\usepackage{dcolumn}
\usepackage{bm}

\begin{document}

\title{Suppression of Faraday waves in a Bose-Einstein condensate in the presence of an optical lattice}

\author{Pablo Capuzzi}
\affiliation{Departamento de Fisica, FCEN
Universidad de Buenos Aires, Ciudad Universitaria, Pab. I
C1428EGA Buenos Aires, Argentina}
\affiliation{Instituto de F{\'\i}sica de Buenos Aires - CONICET, Argentina}
\author{Mario Gattobigio}
\affiliation{Universit\'e de Nice - Sophia Antipolis, Institut non Lin\'eaire de Nice, CNRS, 1361 route des Lucioles, 06560 Valbonne, France}

\author{Patrizia Vignolo}
\affiliation{Universit\'e de Nice - Sophia Antipolis, Institut non Lin\'eaire de Nice, CNRS, 1361 route des Lucioles, 06560 Valbonne, France}

\begin{abstract}
  We study the formation of Faraday waves in an elongated
  Bose-Einstein condensate in presence of a one-dimensional
  optical lattice, where phonons are parametrically excited by
  modulating the radial confinement of the condensate. For
  very shallow optical lattices, phonons with a well-defined wave
  vector propagate along the condensate, as in the absence of the
  lattice, and we observe the formation of a Faraday pattern. By
  increasing the potential depth, the local sound velocity decreases
  and when it equals the condensate local phase velocity, the condensate
  becomes dynamically unstable and the parametric excitation of
  Faraday waves is suppressed.
\end{abstract}

\maketitle
\section{Introduction}
The term Faraday waves refers to a surface density modulation
generated by the interference of counterpropagating phonons excited by
an oscillatory motion of a nonlinear medium \cite{Faraday1831}. The
formation of Faraday waves has been studied in several physical
contexts \cite{Cross1993}, including convective fluids, nematic liquid
crystals, nonlinear optics, biology, and, recently, in
Bose-Einstein condensates (BECs) \cite{Engels2007}.  In the context of
ultracold gases, Faraday patterns can be excited by the modulation of
the scattering length \cite{Staliunas2002} or by varying the
transverse trap confinement
\cite{Kagan2007,Engels2007,Nicolin2007,Capuzzi2008}, the main
ingredient being the modulation of the nonlinear interaction.  The
study of the formation of spatial structures in a quantum fluid
confined in a smooth potential, such as the cigar-shaped harmonic trap
used in the BEC experiment \cite{Engels2007}, can give access to
information about the excitation spectrum (and thus about the equation
of state of the system) through the relation between the excitation
frequency and the measured wave vector of the Faraday wave.  This is
somehow similar to Bragg spectroscopy on ultracold gases
confined in an optical lattice, where the excitation spectrum is
inferred by exposing the system to a periodic lattice modulation and
measuring the energy absorption
\cite{Stoeferle2004,Ozeri2005,Tozzo2005,Kollath2006,Fallani2007}.
Proceeding along this line, one may expect to probe the spectrum of an
ultracold gas, {\it even in the presence of an optical lattice}
\cite{Kramer2005}, by observing the formation (or the lack of
formation) of a Faraday pattern for {\it any} lattice depth provided
that the system remains compressible (i.e. the lattice does not induces a
superfluid-Mott transition).

In this paper we show that the presence of an optical lattice may
dramatically alter the parametric excitation of Faraday waves. In
particular, we show that the formation of the Faraday waves can be
suppressed not only by the appearance of a gap in the spectrum of
collective excitations, but also by the fact that the local flow of the 
condensate
exceeds the Landau critical velocity \cite{Watanabe2009}. We focus our
study on a cigar-shaped condensate where phonons are parametrically
excited by modulating the tight radial confinement at a frequency
$\Omega$. The parametrically excited phonons have an energy
$\hbar\Omega/2$ and a Faraday wavevector $q_{F}=q(\Omega/2)$
determined by the energy spectrum of the system. We add a
one-dimensional (1D) optical lattice along the axial direction, and,
by increasing the lattice potential depth $V_0$, we observe the
suppression of the parametric excitation of phonons with wavevector
$q_F$, even if $\hbar\Omega/2$ is still an allowed energy of the
collective modes.  The sound mode at $q_F$, excited by the parametric
instability, is submerged by several other modes excited by the
dynamical instability.  This instability can be related to the
previously-studied instabilities in bosonic superfluid currents moving 
with respect to an optical lattice potential. In these set-ups the current
remains stable if the superflow momentum does not exceed half the
recoil momentum \cite{Sarlo2005,Mun}.  At variance, as in our system
the condensate center-of mass is at rest, the dynamical instability
washing out the Faraday pattern is due to a {\it local} supersonic
flow through the periodic potential induced by the continuous driving
of the radial breathing mode.  Indeed, we observe that the critical
value $V_{0,{\rm crit}}$ at which Faraday waves are suppressed
corresponds to the critical value of the local BEC flow
\cite{Watanabe2009}.

The article is organized as follows. In Sec.~\ref{sec:model} we
introduce the nonlinear Schr{\"o}dinger equations that describes the
condensate in the elongated geometry. We make use of the three-dimensional (3D)
Gross-Pitaevskii equation (GP) for the evaluation of the low-lying
energy spectrum (Sec.~\ref{sec:spectrum}) and of a time-dependent
nonpolynomial nonlinear Schr{\"o}dinger equation (NPSE) to numerically
study the spatial Faraday pattern formation (Sec.~\ref{sec:num}). In
Sec.~\ref{sec:num} we compare the results for a cylindrical and a
cigar-shaped trap and show that the suppression of the parametric
excitations of Faraday waves occurs when the local condensate flow
reaches the critical value associated to the Landau criterion. Our
concluding remarks are given in Sec.~\ref{sec:summ}.

\section{Theoretical framework}
\label{sec:model}
We study a 3D Bose condensate trapped into a cigar-shaped potential,
possibly with a 1D periodic potential superimposed along the axial
direction. The full description of the system in the limit of zero
temperature and low density is given by the 3D GP
\begin{equation}
  i\hbar \frac{\partial}{\partial t} \psi(\mathbf r, t)
  =
  \bigg[-\frac{\hbar^2}{2m}\nabla^2 + U(\mathbf r) + gN|\psi(\mathbf r, t)|^2
  \bigg] \psi(\mathbf r, t) \,,
  \label{GPE3D}
\end{equation}
with $\psi(\mathbf r, t)$ the wavefunction of the condensate, $m$ the
mass of the atoms composing the condensate, $N$ the number of
particles, and $g=4\pi\hbar^2 a_s/m$ the interaction coupling constant
with $a_s$ the $s$-wave scattering length between particles.  In our setup,
the trapping potential $U(\mathbf{r})$ is the sum of a cigar-shaped
harmonic trap plus an optical potential $V(z)$
\begin{equation}
  U(\mathbf r) = \frac{1}{2}m\,\omega_\perp^2(x^2+y^2) + \frac{1}{2} m\,\omega_z^2
  z^2 + V(z)\,,
  \label{}
\end{equation}
with $\omega_\perp$ the trapping frequency in the perpendicular direction, 
$\omega_z$ the trapping frequency in the longitudinal direction, and 
$\omega_z \gg \omega_\perp$.
The additional potential $V(z)$ is either zero or equal to a periodic 
potential $V(z)=V_0\sin^2(q_Bz)$, with periodicity $d=\pi/q_B$.

Given the above trap geometry, it has been shown in
Ref.~\cite{Salasnich2002} that a reliable description of the
condensate dynamics is given by an effective 1D time-dependent
NPSE.  This
equation is derived by a variational ansatz for the wavefunction 
\begin{equation}
  \psi(\mathbf r, t) = \phi(x,y,t,\sigma(z,t))\,f(z,t) \,,
  \label{}
\end{equation}
where the transverse wavefunction $\phi$ is modelled by a Gaussian
function
\begin{equation}
  \phi(x,y,t,\sigma(z,t)) = \frac{1}{\sqrt\pi\sigma(z,t)} 
  \text{e}^{-(x^2+y^2)/2\sigma(z,t)^2}\,,
  \label{}
\end{equation}
with a time- and longitudinal-dependent variance $\sigma(z,t)$.  The
validity of this description is based on the assumption that the
transverse wavefunction $\phi$ is slowly varying in such a way that the radial
velocity can be neglected.  Furthermore, it has been shown that
neglecting the transverse excitations does not affect the prediction
for energy and dynamical instabilities
\cite{Modugno2004}, or the onset of Faraday waves for
a parametrically excited condensate \cite{Nicolin2007}.  Both the
longitudinal wavefunction $f(z,t)$ and the variance $\sigma(z,t)$ are
determined by the variational principle for the energy, and the result
is that the longitudinal wavefunction is governed by the NPSE
\begin{eqnarray}
i\hbar\dfrac{\partial}{\partial t}f&=&
\left[-\dfrac{\hbar^2}{2m}\dfrac{\partial^2}{\partial z^2}+
\frac{1}{2}m\,\omega_z^2 z^2 + V(z)+ \right.\nonumber \\
&&\left.\hbar\omega_\perp\dfrac{1+3a_sN|f|^2}{\sqrt{1+2a_sN|f|^2}}\right]f \,,
\label{nostraequazione}
\end{eqnarray}
while the variance is algebraically determined by $f(z,t)$
\begin{equation}
  \sigma^2(z,t) = \frac{\hbar}{m\,\omega_\perp}
  \sqrt{1+2a_sN\,|f(z,t)|^2} \,.
  \label{}
\end{equation}
The 3D density profile and velocity field can be obtained as 
\begin{eqnarray}
\rho(\mathbf{r})&=&\tilde{\rho}(z)\,\frac{e^{-r^2/\sigma^2}}{\pi\sigma^2}\nonumber \\
\mathbf{v}(\mathbf{r})&=&v(\mathbf{r})\,\hat{z}=\frac{\hbar}{2mi}
\frac{f'^{*}(z)f(z)-f'(z)f^{*}(z)}{\tilde{\rho}}\hat{z},
\end{eqnarray}
with $\tilde{\rho}(z)=|f|^2$ the integrated 1D density.  In the
following we will consider the case of a radial confinement
$\omega_{\perp 0}/2\pi=200$ Hz and lattice spacing $d=4680$nm;
this gives a  recoil energy $E_r=\hbar^2\,q_B^2/2m\simeq 0.14
\hbar\omega_{\perp,0}$. The potential depth $V_0$ will be used as
a parameter. 
\section{The low-lying energy spectrum with periodic potential}
\label{sec:spectrum}
\begin{figure}
\includegraphics[width=0.9\columnwidth,clip=true]{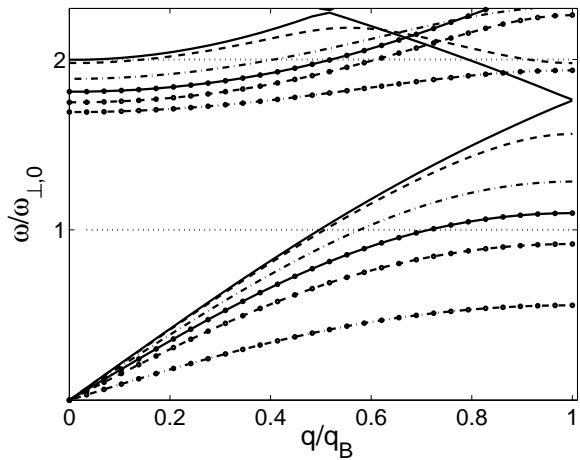}
\caption{\label{fig:band}Low-lying energy spectrum in the hydrodynamic
  approximation for the infinity lattice for $V_0/E_r=$ 0
  (continuous line), 50 (dashed line), 130 (dot-dashed line), 190
  (circles and continuous line), 250 (circles and dashed line), 350
  (circles and dot-dashed line).}
\end{figure}

We calculate the low-lying energy spectrum of the condensate subject
to a cylindrical confinement ($\omega_z=0$) with periodic boundary
conditions (PBC) for different lattice potential depths. Writing
Eq. (\ref{GPE3D}) in terms of the particle density $\rho$ and velocity
${\bf v}$ and considering the hydrodynamic limit of the 3D GP
\cite{Stringari1996}, we calculate the collective modes spectrum
$\omega(q)$ solving the eigenvalue equation
\begin{equation}
-m\omega^2\delta\rho=g\nabla\cdot(\rho_0\nabla\delta\rho),
\label{eq:modes}
\end{equation}
$\rho_0$ being the equilibrium density in the Thomas-Fermi
approximation, and $\delta\rho({\bf r})=\delta\rho(r_\perp, z)e^{iqz}$
the density perturbation with zero angular momentum propagating along
the $z$ axis.  For that matter we discretize Eq. \
(\ref{eq:modes}) in a rectangular domain $(r,z)$, and impose PBC in
$z$, i.e., $\delta\rho(r_{\perp},z+d)=\delta\rho(r_{\perp},z)$.
Figure \ref{fig:band} shows the low-lying energy spectrum for the case
of $N_0=3.2\times 10^5$ particles per site and several values of $V_0$.
As expected, by increasing the lattice depth, the lowest-energy band
gets narrower and an energy gap opens at $q=q_B$.

\section{The Faraday wave excitations}
\label{sec:num}
In this section we present our numerical results for dynamics of the
condensate in the presence of an optical lattice.  To excite Faraday
waves we proceed as follows: first we calculate the ground state
$f_0(z)$ for a static potential, setting $\omega_\perp=\omega_{\perp
  0}$, then we switch on the trap modulation $\omega_{\perp}(t) =
\omega_{\perp 0}(1+\epsilon\cos\Omega t)$ at the frequency
$\Omega$ and amplitude $\epsilon$. We fix
$\epsilon=0.1$ and choose $\Omega=2\omega_{\perp,0}$ in order to be
close enough to the natural breathing mode in absence of the lattice.

We focus on the dynamics of the density and velocity field of the
atoms for increasing values of the lattice depth, and compare the case
of an infinite cylinder, i.e. $\omega_z=0$, and the case of a cigar-shaped BEC,
for the same number $N_0$ of particles in the central site.
The NPSE is numerically solved using a split-step method and spatial
Fast Fourier transforms (FFT) in a finite domain with PBC. In our
numerical calculations, we found that is enough to consider a spatial
grid with 1024 points and a time step $\delta t\simeq
10^{-4}/\omega_{\perp,0}$ to follow the dynamics for long times up to
several hundreds of the transverse frequency period.

\subsection{The infinite cylinder}

To analyze the formation of a pattern on top of the time-modulated
density we proceed as follows.  At each time we separate 
the part with the spatial periodicity of the 
optical lattice from the total density profile.
Explicitely, we write $\tilde\rho(z,t) = \tilde\rho_0(z,t) +
\delta\tilde\rho$ where
\begin{equation}
\tilde\rho_0(z,t) = A(t) + B(t)\cos(2q_B\,z + \varphi(t)),
\label{eq:analisis_rho}
\end{equation}
with $A, B$ and $\varphi$ fitting parameters that characterize the
renormalized trap, and $\delta\tilde\rho$ the remaining
fluctuation. The Faraday pattern will be analyzed in terms of the
behaviour of this fluctuation.

\begin{figure}
\includegraphics[width=0.9\columnwidth,clip=true]{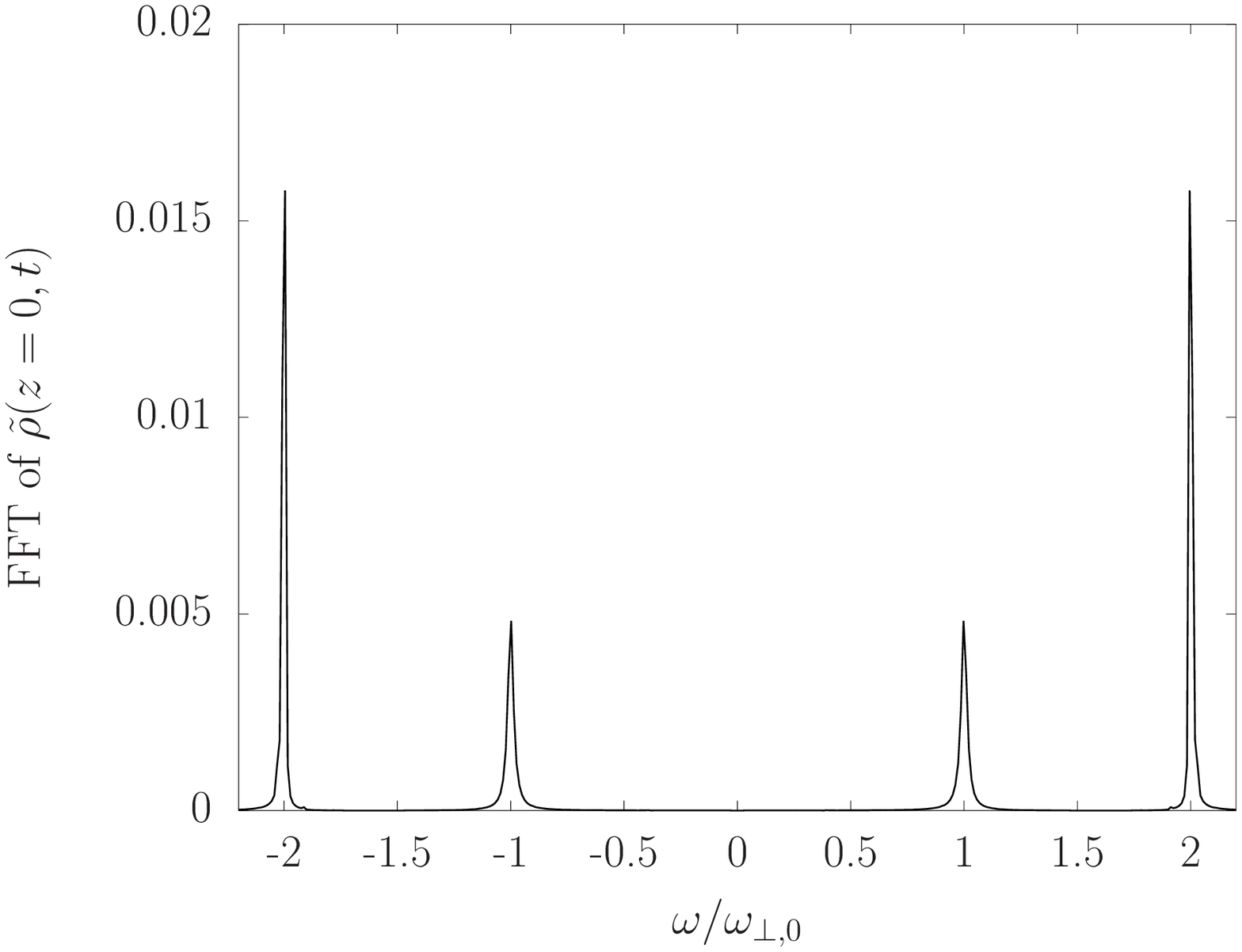}
\caption{\label{fig:central_otro_V0_20}Temporal FFT of the central density from
$\omega_{\perp,0}t=0$ to $530$ with a lattice depth $V_0/E_r=20$.}
\end{figure}

\begin{figure}
\includegraphics[width=0.9\columnwidth,clip=true]{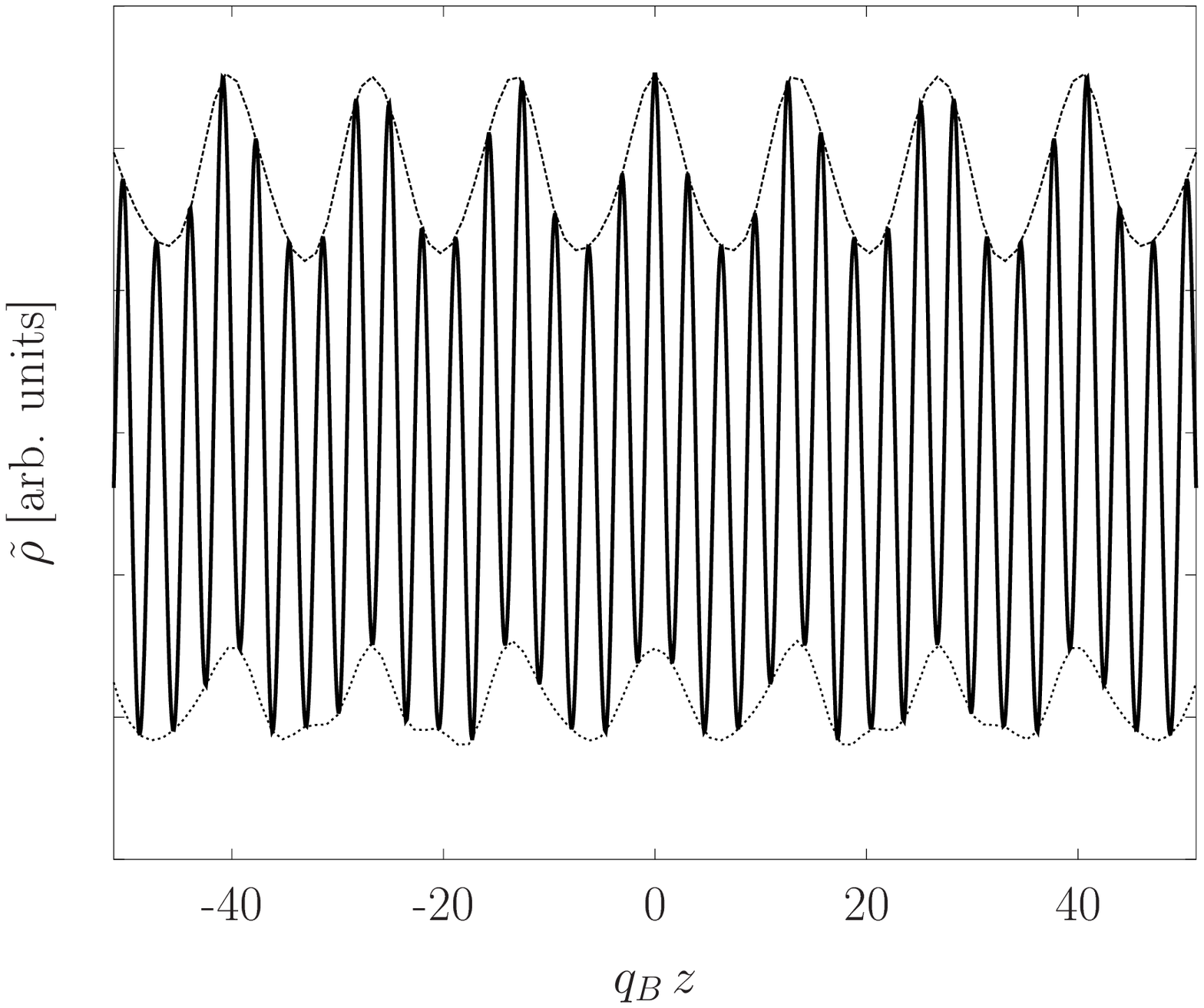}
\includegraphics[width=0.9\columnwidth,clip=true]{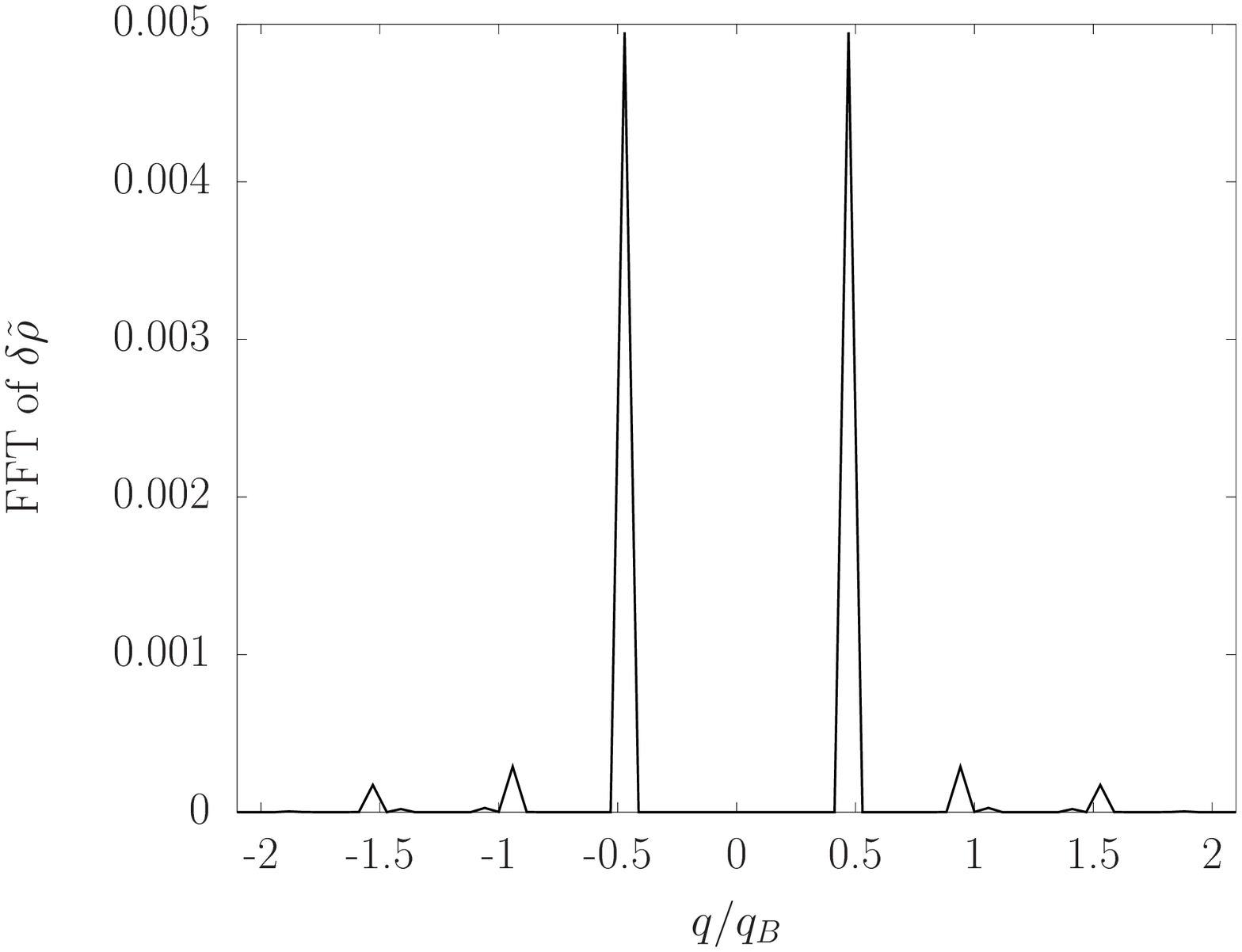}
\caption{\label{fig:denz_otro_V0_20}Faraday pattern in a lattice with
  depth $V_0/E_r=20$. The top panel shows the density profile
  $\delta\rho(z)$ as a function of $q_Bz$ at $\omega_{\perp,0}t=530$.
  The bottom panel shows the spatial FFT of the fluctuation
  $\delta\tilde\rho(z)$ at the same time.}
\end{figure}

In Fig.\ \ref{fig:central_otro_V0_20} we show the Fourier transform of
central density $\tilde{\rho}(z=0)$ for a lattice depth $V_0/E_r=20$.
At $\omega_{\perp,0}\,t\simeq 300$ the oscillation starts departing from
a simple scaling solution and a different mode gets populated. This is
seen clearly in the temporal FFT of the central density $\tilde\rho
(z=0)$ depicted in Fig.\ \ref{fig:central_otro_V0_20} where the
frequency $\omega=\Omega/2=\omega_{\perp,0}$ shows up. Shortly after,
a well defined Faraday pattern is superimposed on the lattice.  This
is shown in Fig.\ \ref{fig:denz_otro_V0_20} where the linear density
$\tilde\rho(z)$ and the spatial FFT of its fluctuation
$\delta\tilde\rho(z)$ are plotted at a time $\omega_{\perp,0}t=530$.  It is
worthwhile noticing that in the spatial FFT of $\delta\tilde\rho$ the
peaks at $q=\pm2q_B$ are never present because of the choice of the
fitting function $\tilde\rho_0$ in Eq. (\ref{eq:analisis_rho}).  The
peaks in the bottom panel at $q=\pm 0.49q_B$ are the wavevectors of
the counter-propagating phonons giving rise to the Faraday pattern, as
expected from Fig. \ref{fig:band}.  For longer times many more modes
are populated and a pure Faraday excitation cannot be observed.

\begin{figure}
\includegraphics[width=0.9\columnwidth,clip=true]{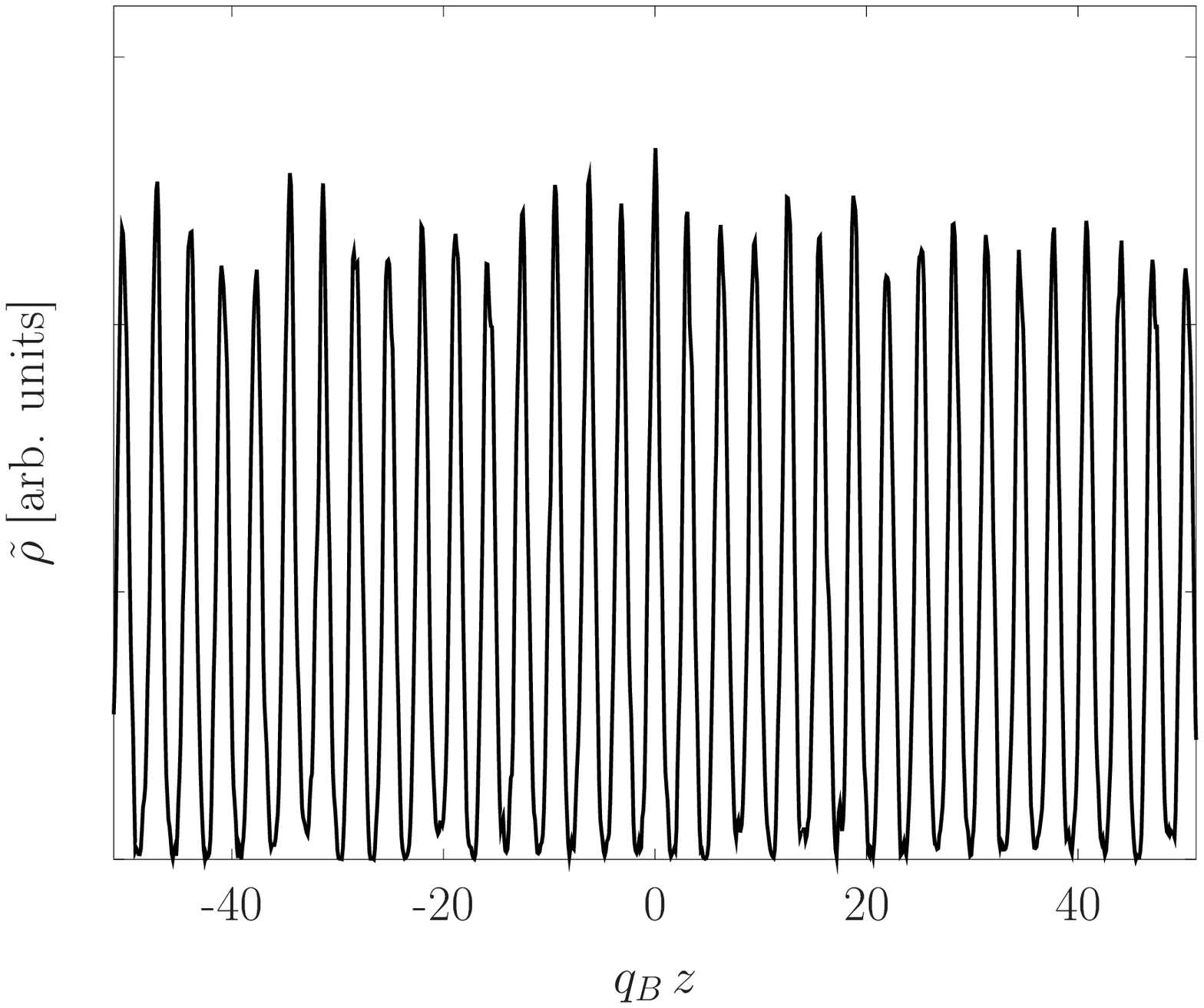}
\includegraphics[width=0.9\columnwidth,clip=true]{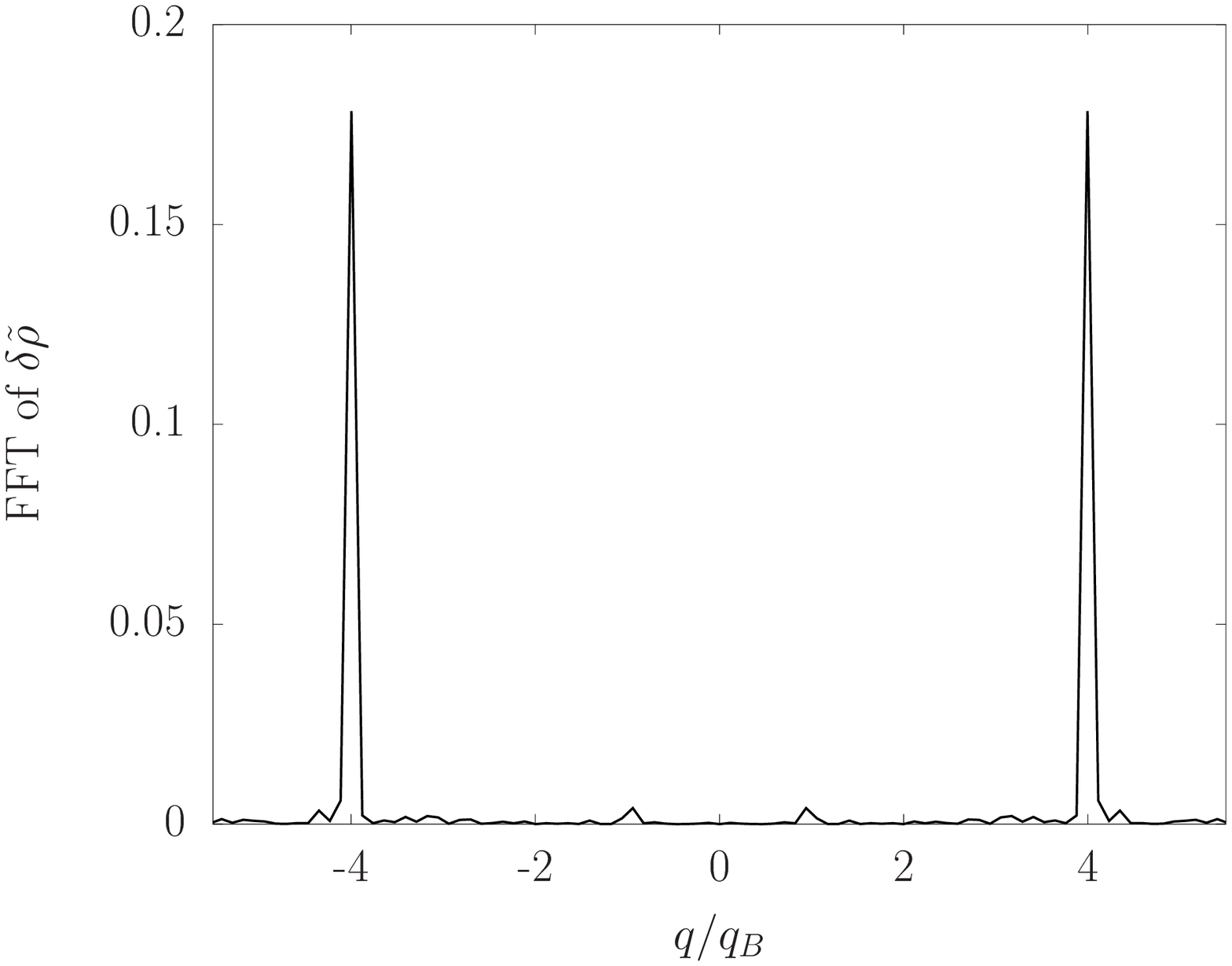}
\caption{\label{fig:densz_otro_V0_350}
The same as in Fig.\ \ref{fig:denz_otro_V0_20}, but for
$V_0/E_r=350$ and $\omega_{\perp,0}t=150$. In this case we
observe the contribution from higher Brillouin zones}
\end{figure}

For deep lattices ($V_0/E_r\gtrsim 230$), the frequency $\Omega/2$
lies within the energy gap and no Faraday pattern can be
formed. However, as $\Omega$ can intersect an energy band, we may
expect fluctuations associated to the wavelength of the
corresponding collective excitation. In particular, at $V_0/E_r=350$
modes at $\Omega/2$ are
not excited anymore, while modes at $\omega=\Omega$ are still
present. In the spatial domain (see Fig. \ref{fig:densz_otro_V0_350})
two main features can be detected: i) contributions from higher Brillouin
zones at $q=\pm2q_B\,n$, stemming from the periodicity of the GS
density profile which is not contained in the weak $V_0$ limit ansatz for
$\tilde{\rho}_0(z,t)$ in Eq.\ (\ref{eq:analisis_rho}); ii) a weak mode
in the vicinity of $q_B$, originating from the collective mode at
$\omega = \Omega$ (cf. the corresponding upper-band in the Fig. \ref{fig:band}).

\begin{figure}
\includegraphics[width=0.9\columnwidth,clip=true]{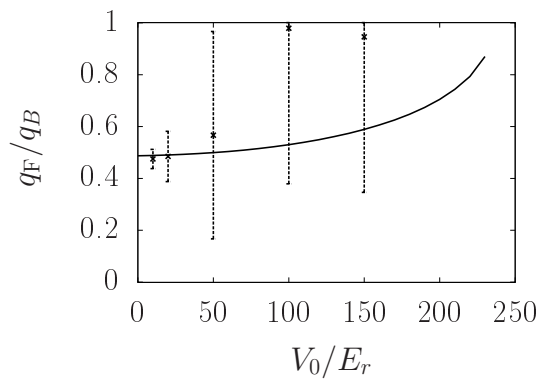}
\caption{\label{fig:sumkPBC}Main wavevector $\langle
|q|\rangle_{\text{min}}$ (stars) identified in the density fluctuation
of the lattice during a time interval
$\omega_{\perp,0}t\simeq1500$ and its dispersion $\Delta q$
(schematized as error bars) compared with the prediction of a
Faraday pattern with a wavelength $q_F$ (solid line), as deduced
by the energy spectrum evaluated in the hydrodynamic approach.}
\end{figure}

Although for intermediate values of $V_0$, well before
approaching the edge of the first Brillouin zone, one naturally
expects to excite the pattern, we found that for $V_0/E_r> 50$,
the mode at $\omega=\Omega/2$ is not seen to increase
substantially, and therefore it is not possible to identify a
Faraday pattern.  Indeed, we observe several excited wavevectors.
In order to quantify the spreading in the $q$-space as a function
of $V_0$ we calculated the minimum of the average of $|q|$,
$\langle |q|\rangle_{\text{min}}$ over the excitation
$\delta\tilde{\rho}$ during a given time interval. This is
displayed is Fig.\ \ref{fig:sumkPBC} together with its dispersion
$\Delta q=(\langle q^2\rangle -\langle |q|\rangle ^2)^{1/2}$
(schematized as error bars in the same figure) evaluated at the
same time as $\langle|q|\rangle_{\text{min}}$.  For $V_0/E_r>
50$ the dispersion  $\Delta q$ is of the same order of $q_B$, more
precisely, roughly all $q$-vectors greater than 0.2-0.3 $q_B$ are excited.
In Fig.\ref{fig:sumkPBC} we compare these results with the
expected $q_F$-value (solid line), as deduced by the low-lying
energy spectrum (see Fig. \ref{fig:band}) setting
$q_F=q(\Omega/2)$. As already anticipated, we can conclude that a
clear Faraday pattern cannot be identified if $V_0/E_r> 50$,
rather before the appearance of the energy gap. 

\subsubsection{The role of the velocity field}
\begin{figure}
\includegraphics[width=\linewidth,clip=true]{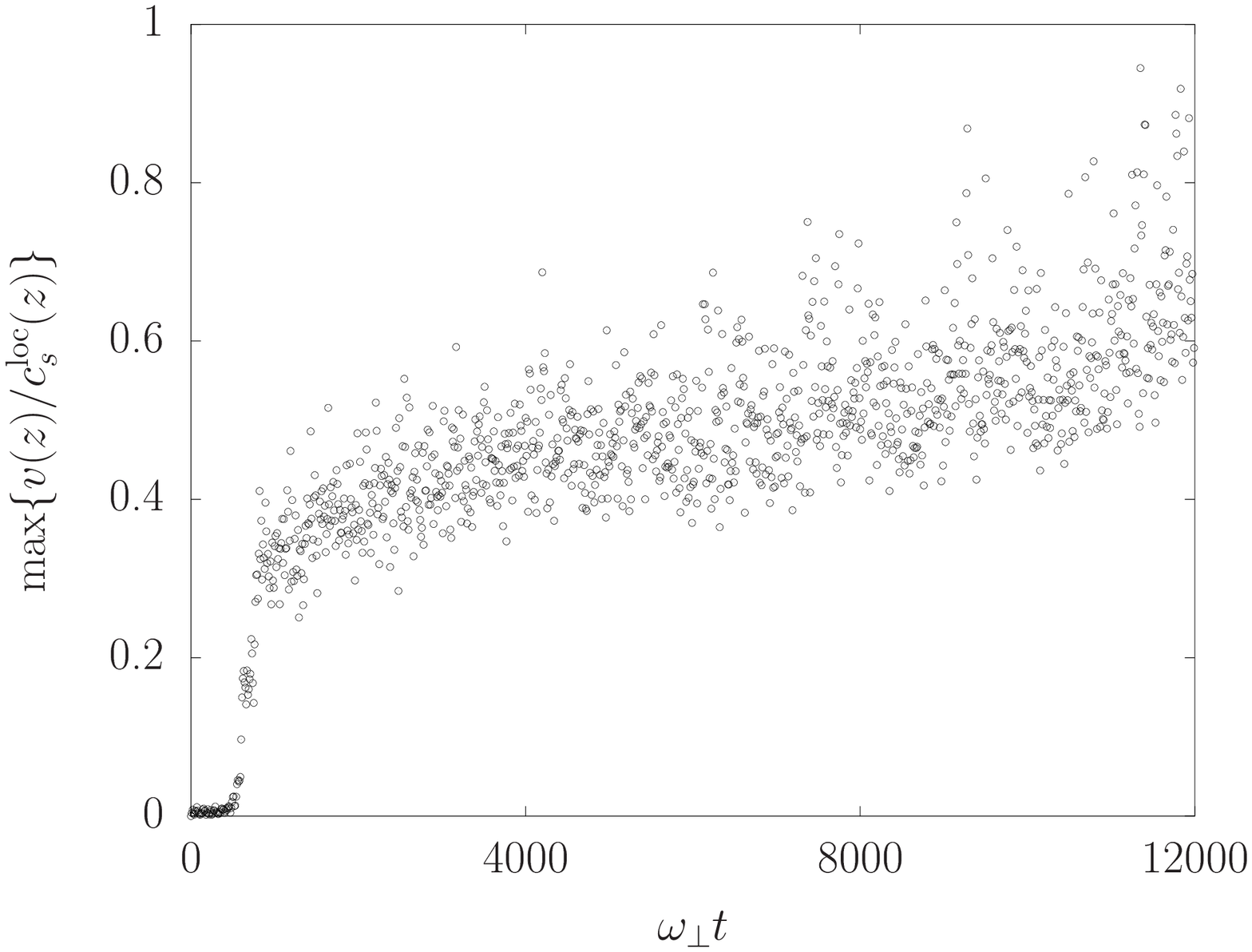} \\
\includegraphics[width=\linewidth,clip=true]{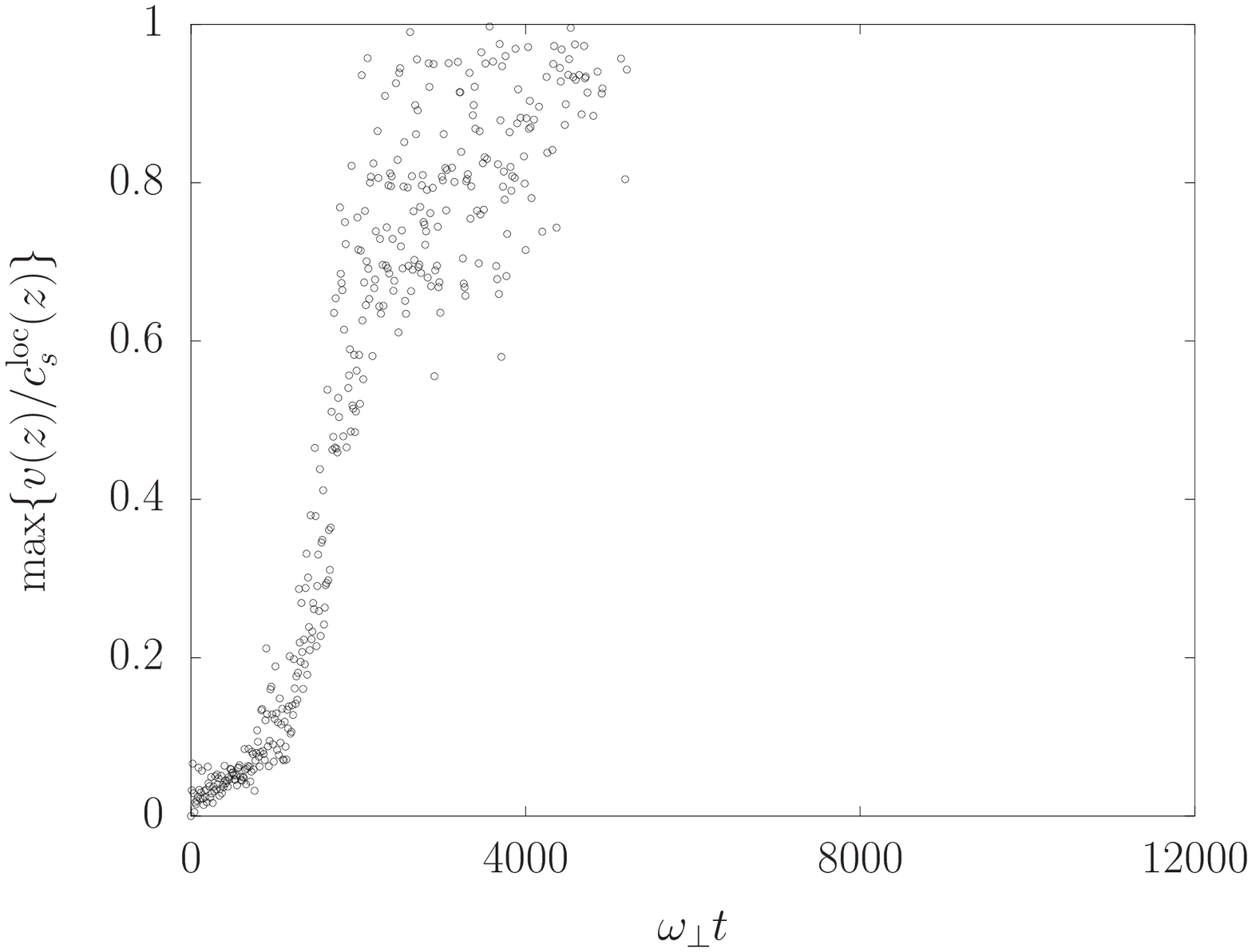}
\caption{\label{fig:sound}
  Maximum of the ratio of the local velocity and the local sound velocity
  for an homogeneous BEC attained over the lattice as functions of
  time for $V_0/E_r=20$ (top panel) and $100$ (bottom panel).}
\end{figure}

Aiming at understanding the mechanism of suppression of the Faraday
wave excitation, we compare the local condensate velocity $v(z)$ with
the local sound velocity $c_s^{\text{loc}}=\sqrt{g\rho/m}$.  At the
maxima of the lattice potential, the central density
$\rho=\tilde\rho/(\pi\sigma^2)$ is minimum, and we expect the ratio
$v(z)/c_s^{\text{loc}}(z)$ takes its maximum value.  Figure \
\ref{fig:sound} shows the behaviour of the maximum value of the ratio
$v(z)/c_s^{\text{loc}}(z)$ as a function of time, during the parametric
excitation, for the case of a shallow (top panel) and of a deep
(bottom panel) lattice.  In the first case, where $v(z)$ is always
lower than $c_s^{\text{loc}}(z)$, the increase of $v(z)$ corresponds
to the onset of the parametric excitation of Faraday waves, while in
the second plot, as soon as $v(z)$ becomes greater than
$c_s^{\text{loc}}(z)$ the superfluid becomes unstable according to the
Landau criterion, and we cannot identify a well-defined Faraday
pattern but several unstable spatial modes. These findings are
qualitatively reproduced if one compares the maximum local velocity to
the sound velocity obtained from the slope of the dispersion
relationship $\omega(q)$.

Since the velocity of the excitations are proportional to the
modulation amplitude $\varepsilon$, one expects to excite the
parametric instability without breaking the superfluid if
$\varepsilon$ is sufficiently small. However this may provoke a
considerable delay for the onset of the Faraday pattern, in both 
numerical and real experiments. Indeed we have investigated this case
for $\varepsilon=0.03$ and found that for $V_0/E_r\ge 100$ we do not
observe the excitation of the Faraday mode during the modulation up to
$\omega_{\perp,0} t = 4.5\times 10^4$, corresponding to roughly 30s.
Since the lifetime of a condensate in most experiments does not exceed
a few seconds, a sizable value of $\varepsilon$, as that chosen in our
first numerical study would be needed to observe the Faraday mode.
In turn, this would give rise to a fast increase of the local flow velocity.  
\subsection{Excitations in a confined lattice}
To allow for a more direct application to current experiments carried
out in optical lattices, hereafter we consider the effects of the
longitudinal harmonic confinement. For that goal we take
$\omega_z/(2\pi)= 20\,$Hz and total number of atoms $N=4.7\times10^6$,
parameters that ensure the same number of atoms $N_0$ 
in the central lattice
site as in the cylindrical confinement.

\begin{figure}
\begin{center}
\includegraphics[width=0.9\columnwidth,clip=true]{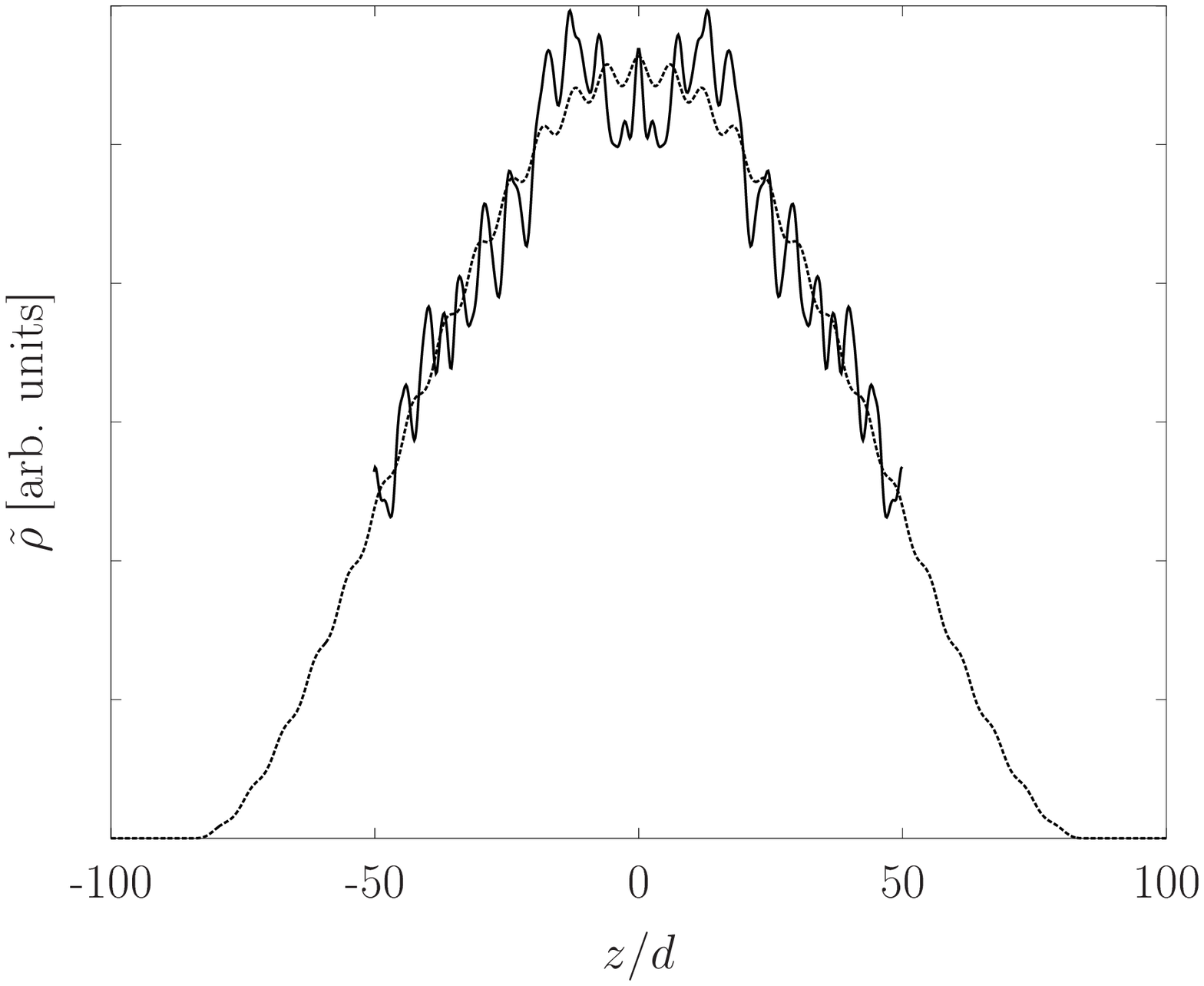}\\
\includegraphics[width=0.9\columnwidth,clip=true]{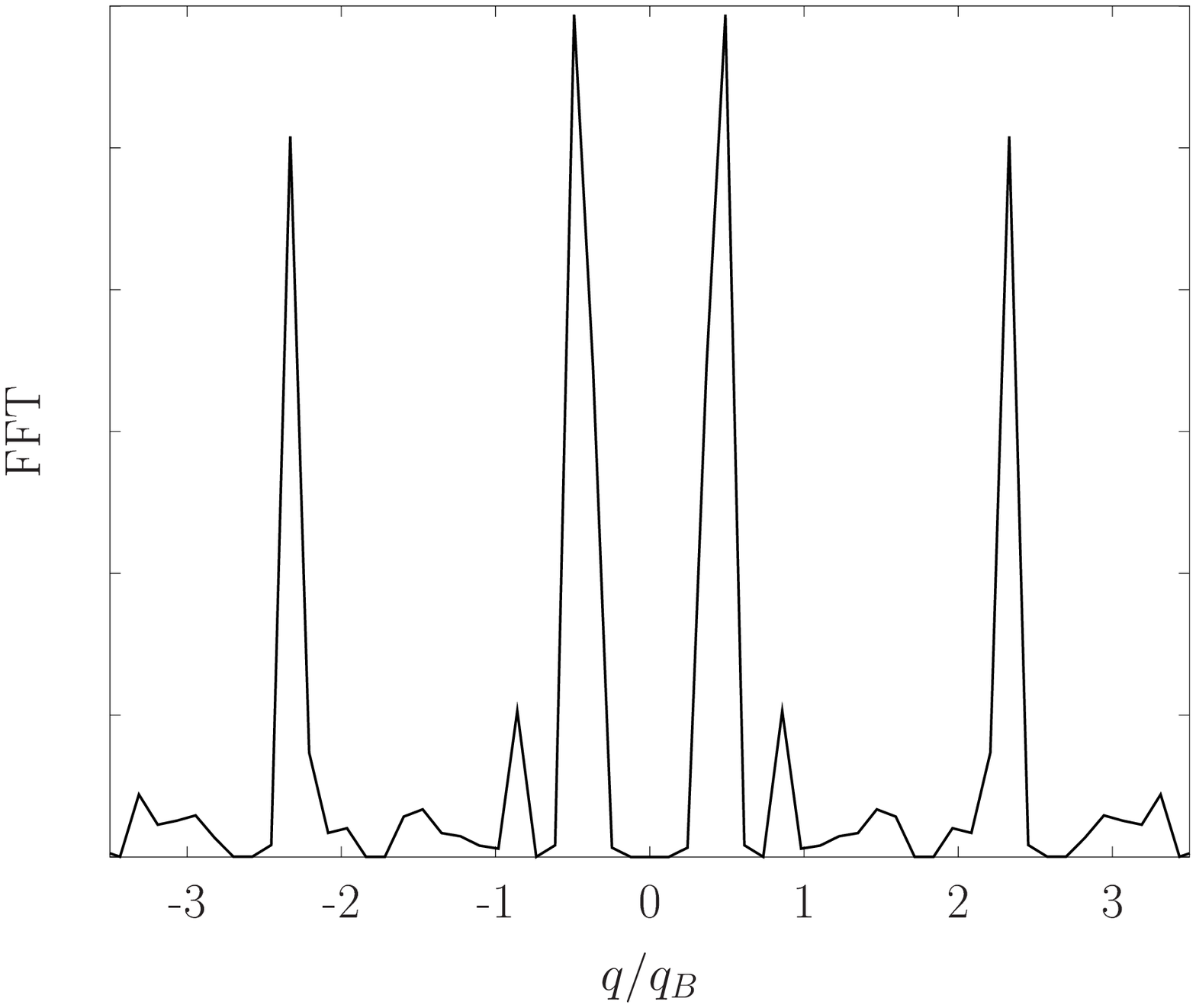}
\end{center}
\caption{\label{fig:conf_res_1}Faraday pattern in an optical lattice
  with $V_0/E_r=1$ confined in an elongated trap. The top panel shows
  the ``zoomed'' density profile $\tilde\rho_0 + 5 \delta\tilde\rho$ at
  $\omega_{\perp,0}t=450$ (solid line) together with the ground-state
  density profile (dashed line). Bottom panel shows the spatial FFT of
  the fluctuation $\delta\tilde\rho$ at the same time.}
\end{figure}

In Fig. \ref{fig:conf_res_1} we show results for $V_0/E_r=1$.  The
continuous line, that corresponds to $\tilde\rho_0 + 5 
\delta\tilde\rho$, the factor of 5 having been chosen to zoom the
density fluctuation, shows the set up of the Faraday pattern at
$\omega_{\perp,0}t=450$ compared to the ground state at $t=0$ (dashed
line).  The presence of the Faraday excitation can be confirmed by the
inspection of the spatial FFT of the density fluctuation at $z=0$.
The lowest $|q|$ peaks correspond to $q_F=q(\Omega/2)$ (see
Fig. \ref{fig:band}), while the second main peaks correspond to
$|q|=q_F+2q_B$, following the lattice periodicity.  The broadening of
the peaks with respect to the spatial FFT in the cylindrical
confinement are due to the inhomogeneous density profile.

Our analysis for the case of a cigar-shaped confinement  is in agreement
with our study for the cylindrical trap, namely the axial confinement,
in the limit where $\omega_z/\omega_{\perp 0}\ll 1$, does not affect
the results shown in the previous section.

\section{Final remarks}
\label{sec:summ}
The spectrum of a bosonic gas in an optical lattice strongly
depends on the lattice potential depth. In the strongly
correlated regime, where there are few particles per well, a deep
lattice potential induces a quantum phase transition towards the
incompressible Mott state.  Otherwise, deep in the superfluid
regime where the average well occupation is large, by increasing
the lattice potential depth the Bogoliubov bands get narrower and
the gaps widen out, as shown in Fig. \ref{fig:band}.  The
parametric excitation of phonons and the observation of the onset
of Faraday waves is expected to be possible if the excited mode
is not in the Mott or Bogoliubov gaps \cite{Kramer2005}.

In this work we point out another mechanism of suppression of Faraday
waves.  During the parametric excitation, there is a local
superfluid flow through the lattice barriers. The presence of
these barriers do not affect the growth of phononic modes with a
well defined wave vector while  the flow velocity remains below
the sound velocity \cite{Watanabe2009}.  By increasing the
potential depth, this condition cannot be fulfilled anymore: the
density at the potential maxima decreases and thus the local sound
velocity decreases too. The superfluid becomes 
unstable and the spectrum of the system cannot be inferred by the
parametric excitation.

\section*{Acknowledgements}
This work was supported by the CNRS-CONICET international cooperation
grant n. 22966.

\end{document}